\documentclass[12pt]{article}
\def\ds{\displaystyle}
\def\l{\lambda}
\newcommand{\bq}{\begin{equation}}
\newcommand{\eq}{\end{equation}}
\newcommand{\ba}{\begin{eqnarray}}
\newcommand{\ea}{\end{eqnarray}}

\newcommand{\nl }{ \nonumber  }

\newcommand{\p}{\partial}
\newcommand{\h}{\hspace{.5cm}}

\newcommand{\la}{\lambda}
\newcommand{\La}{\Lambda}

\begin{document}
\vspace*{1.cm}
\begin{center}
{\bf MAGNON-LIKE DISPERSION RELATION FROM M-THEORY}

\vspace*{1cm} P. Bozhilov${}^{\star}$ and R.C.
Rashkov${}^{\dagger}$\footnote{e-mail: rash@phys.uni-sofia.bg;
bozhilov@inrne.bas.bg}

\ \\
${}^{\star}$ \textit{Institute for Nuclear Research and Nuclear
Energy, Bulgarian Academy of Sciences, \\ 1784 Sofia, Bulgaria}

${}^{\dagger}$ \textit{Department of Physics, Sofia University,
1164 Sofia, Bulgaria}

\end{center}
\vspace*{.5cm}
\begin{abstract}
We investigate  classical rotating membranes in two different
backgrounds. First, we obtain membrane solution in $AdS_4\times
S^7$ background, analogous to the solution obtained by Hofman and
Maldacena in the case of string theory. We find a magnon type
dispersion relation similar to that of Hofman and Maldacena and to
the one found by Dorey for the two spin case. In the appendix of
the paper, we consider membrane solutions in $AdS_7\times S^4$,
which give new relations between the conserved charges.
\end{abstract}


\vspace*{.5cm} {\bf Keywords:} M-theory, AdS-CFT correspondence,
spin chains.

\section{Introduction}

The main directions of developments in String/M theory last years
are related to their relations to the gauge theories at strong
(weak) coupling. A powerful tool in searching for string/M theory
description of gauge theories is AdS/CFT correspondence. One of
the predictions of the correspondence is the equivalence between
the spectrum of free string/M theory on $AdS$ spaces and the
spectrum of anomalous dimensions of gauge invariant operators in
the planar $\mathcal{N}=4$ Supersymmetric Yang-Mills (SYM) theory.
Since the string/M theory in such spaces is highly non-linear, the
check of this conjecture turns out to be very nontrivial. The
known tests that confirm the correspondence beyond the
supergraviry approximation are based on the suggestion by Gubser,
Klebanov and Polyakov \cite{gkp}, that one can look for certain
limits where semiclassical approximation takes place and the
problem becomes tractable and some comparisons on both sides of
the correspondence can be made. From string/M theory point of view
this means that one should consider solutions with large quantum
numbers, which are related to the anomalous dimensions of gauge
theory operators. While to find the spectrum from this side,
although complicated, is possible, a reliable method to do it from
gauge theory side was needed. Minahan and Zarembo proposed a
remarkable solution to this problem \cite{mz} by relating the
Hamiltonian of the Heisenberg spin chain with the dilatation
operator of $\mathcal{N}=4$ SYM. On other hand, in several papers,
the relation between strings and spin chains was established, see
for instance \cite{kruczenski},\cite{dim-rash},\cite{
lopez},\cite{tseytlin1} and references therein. This idea opened
the way for a remarkable interplay between spin chains, gauge
theories, string theory \footnote{For very nice reviews on the
subject with a complete list of references see
\cite{beisphd},\cite{tseytrev},\cite{Plefka:2005bk}} and
integrability (the integrability of classical strings on
$AdS_5\times S^5$ was proven in \cite{bpr}). One of the ways to
compare these two sides of the AdS/CFT correspondence is to look
for string/M theory solutions corresponding to different corners
of the spectrum of the spin chains arising from string and gauge
theory sides. Although there is no known direct relation of
M-theory in the limit of large quantum numbers to spin chains, one
can still directly relate the dispersion relations obtained from
M-theory to the spectrum of gauge spin chains.

The most studied cases were spin waves in long-wave approximation,
corresponding to rotating and pulsating strings in certain limits,
see for instance the reviews
\cite{beisphd},\cite{tseytrev},\cite{Plefka:2005bk} and references
therein. Another interesting case are the low lying spin chain
states corresponding to the magnon excitations. One class of
string/membrane solutions already presented in a number of papers
is the string/M theory on pp-wave backgrounds. The later, although
interesting and important, describe point-like objects which are
only part of the whole picture. The question of more general
string/membrane solutions corresponding to this part of the
spectrum was still unsolved.

Recently Maldacena and Hofman\cite{HM} were able to map spin chain
"magnon" states to specific rotating semiclassical string states
on $R\times S^2$. This result was soon generalized to magnon bound
states (\cite{Dorey1},\cite{Dorey2},\cite{AFZ},\cite{MTT}\cite{MRT}), dual
to strings on $R\times S^3$ and $R\times S^5$ with two and three non-vanishing angular
momenta.
The relation between energy and angular momentum for the one spin
giant magnon found in \cite{HM} is:

\begin{equation}
 E-J=\ds\frac{\sqrt{\lambda}}{\pi}\mid\sin\ds\frac{p}{2}\mid,
\end{equation}
where $p$ is the magnon momentum, which on the string side is
interpreted as a difference in the angle $\phi$ (see \cite{HM} for
details). In the two spin case, the $E-J$ relation was found both
on the string \cite{Dorey2},\cite{AFZ},\cite{MTT} and spin chain
\cite{Dorey1} sides and looks like:

\begin{equation}
 E-J=\ds\sqrt{J_2^2+\ds\frac{\l}{\pi^2}\sin^2\frac{p}{2}},
\end{equation}
where $J_2$ is the second spin of the string.

In this paper, we are looking for membrane solutions analogous to
giant magnon strings. Due to the AdS/CFT correspondence, the
dispersion relations are expected to give similar result to that
in the case of string magnon states. Indeed, for one particular
ansatz for the embedding coordinates, we find dispersion relation
that is similar to those obtained from strings for the magnon part
of the spectrum of the gauge theory spin chain.
One may wonder about how general this solution is. Certainly there
are more examples of such solutions, which we will report in the near future,
rising the conjecture that this kind of dispersion relations captures an essential feature of membrane spectrum.
 Our result give a
support to the M/gauge theory correspondence previously checked
for particular parts of the spectrum
\cite{6}-\cite{B0603}\footnote{See also \cite{MS1}, \cite{MS2} and
\cite{AFP02}.}. Apart from this result, we present in the Appendix
different solutions for embeddings in $AdS\times S^1$ part of the
geometry. The results give different complicated dispersion
relations for which, unfortunately, we don't have conclusive
interpretation from gauge theory side. We hope that these
solutions might be also useful for establishing M/gauge theory
duality in the future.

\setcounter{equation}{0}
\section{Giant magnons from M-theory}

The idea we will follow in this section is inspired from the
analogy with the string theory case. In the later case the magnon
dispersion relations are found by considering dynamics of an
arc-like string with two ends on the equator of the five sphere
$S^5$, or spiky strings. To capture the essence of this dynamics
it is important to consider the correct embedding of the string
configurations. We will look for an embedding which is arc-like
but extended on the second spacial coordinate of the membrane. It
can be though as a continuous  family of arcs parameterized by the
spacial coordinate  $\xi^2$. This analogy will allow us latter on
to make reduction and compare with the string case. We will work
with the following gauge fixed membrane action and constraints
\cite{B0603}, which {\it coincide} with the usually used gauge
fixed Polyakov type action and constraints after the
identification \cite{27} $2\lambda^0T_2=L=const$: \ba\label{omagf}
&&S_{gf}= \int
d^{3}\xi\left\{\frac{1}{4\lambda^0}\Bigl[G_{00}-\left(2\lambda^0T_2\right)^2\det
G_{ij}\Bigr] + T_2 C_{012}\right\},
\\ \label{00gf} &&G_{00}+\left(2\lambda^0T_2\right)^2\det G_{ij}=0,
\\ \label{0igf} &&G_{0i}=0.\ea In (\ref{omagf})-(\ref{0igf}), the fields
induced on the membrane worldvolume $G_{mn}$ and $C_{012}$ are
given by \ba\label{im} G_{mn}= g_{MN}\p_m X^M\p_n X^N,\h C_{012}=
c_{MNP}\p_{0}X^{M}\p_{1}X^{N}\p_{2}X^{P}, \\ \nl
\p_m=\p/\p\xi^m,\h m = (0,i) = (0,1,2),\h M = (0,1,\ldots,10),\ea
where $g_{MN}$ and $c_{MNP}$ are the target space metric and
3-form gauge field respectively. The equations of motion for
$X^M$, following from (\ref{omagf}), are as follows
$(\mathbf{G}\equiv\det{G_{ij}})$ \ba\label{eqm}
&&g_{LN}\left[\p_0^2 X^N - \left(2\lambda^0T_2\right)^2
\p_i\left(\mathbf{G}G^{ij}\p_j X^N\right)\right]\\ \nl
&&+\Gamma_{L,MN}\left[\p_0 X^M \p_0 X^N -
\left(2\lambda^0T_2\right)^2 \mathbf{G}G^{ij}\p_i X^M \p_j
X^N\right]\\ \nl &&=2\la^0 T_2
H_{LMNP}\p_0X^{M}\p_1X^N\p_2X^{P},\ea where \ba\nl
\Gamma_{L,MN}=g_{LK}\Gamma^K_{MN}=\frac{1}{2}\left(\p_Mg_{NL}
+\p_Ng_{ML}-\p_Lg_{MN}\right)\ea are the components of the
symmetric connection corresponding to the metric $g_{MN}$ and
$H_{LMNP}$ is the field strength of the $3$-form field $c_{MNP}$.

Here, we will search for rotating M2-brane configuration, which
eventually could reproduce the string theory and spin chain (field
theory) results for the two spin giant magnons
\cite{Dorey1}-\cite{MTT}. Namely, we are interested in deriving an
energy charge relation of the type \ba\nl E-J_2=\sqrt{J_1^2 +
\frac{\lambda}{\pi^2}\sin^2 \frac{p}{2}}\ea for \ba\nl
E\to\infty,\h J_2\to\infty,\h E-J_2 - \mbox{finite},\h J_1 -
\mbox{finite}.\ea This relation has been established for strings
on $R\times S^3$ \cite{Dorey2}-\cite{MTT} and $AdS_3\times S^1$
\cite{MTT} subspaces of $AdS_5\times S^5$. Such subspaces also
exist in M-theory backgrounds $AdS_4\times S^7$ and $AdS_7\times
S^4$. However, we have not been able to find rotating M2-brane
configurations with the desired properties on these subspaces of the target spaces.
The experience in these computations led us to the conclusion that
in order the membrane to have the needed semiclassical behavior,
it must be embedded in a space with at least one dimension higher.
In the case of $AdS_7\times S^4$, the task is more difficult to
solve, because of the presence of nontrivial 3-form background
gauge field for the subspace $R\times S^4$. Hence, to simplify the problem
we choose to
consider M2-brane moving on the following subspace of $AdS_4\times
S^7$ \ba\nl ds^2=(2l_pR)^2\left\{-dt^2+4\left[d\psi^2+\cos^2\psi
d\varphi_1^2+\sin^2\psi\left(\cos^2\theta_0
d\varphi_2^2+\sin^2\theta_0d\varphi_3^2\right)\right]\right\},\ea
where the angle $\theta$ is fixed to an arbitrary value
$\theta_0$, and for which the background 3-form field on $AdS_4$
vanishes.

We start with the following ansatz for the membrane \ba\nl
&&X^0(\xi^m)\equiv t(\xi^m)= \La_0^0\xi^0, \h
X^1(\xi^m)=\psi(\xi^2),
\\ \nl &&X^2(\xi^m)\equiv \varphi_1(\xi^m)=
\La_0^2\xi^0,
\\ \nl &&X^3(\xi^m)\equiv \varphi_2(\xi^m)=
\La_0^3\xi^0,
\\ \nl &&X^4(\xi^m)=\varphi_3(\xi^m)=
\La_i^4\xi^i,\h \La_0^0,\ldots,\La_i^4=constants.\ea It
corresponds to M2-brane extended in the $\psi$- direction, moving
with constant energy $E$ along the $t$-coordinate, rotating in the
planes defined by the angles $\varphi_1$, $\varphi_2$, with
constant angular momenta $J_1$, $J_2$, and wrapped along
$\varphi_3$.

It is easy to check that for this choice of the membrane
embedding, the constraints (\ref{0igf}) are identically satisfied.
The remaining constraint (\ref{00gf}), which for the case under
consideration is first integral of the equation of motion for
$\psi(\xi^2)$ following from (\ref{eqm}), takes the form
\ba\label{ecp} &&K(\psi)\psi'^2+V(\psi)=0,\\ \nl
&&K(\psi)=-2^{10}(l_pR)^4(\lambda^0T_2\La_1^4\sin\theta_0)^2\sin^2\psi,
\\ \nl &&V(\psi)=(2l_pR)^2\left\{(\Lambda_0^0)^2-4(\Lambda_0^2)^2
-4\left[(\Lambda_0^3)^2\cos^2\theta_0-(\Lambda_0^2)^2\right]\sin^2\psi\right\}.\ea
From (\ref{ecp}) one obtains the turning point ($\psi'=0$) for the
effective one dimensional motion \ba\label{M^2}
M^2=\frac{(\Lambda_0^0)^2-4(\Lambda_0^2)^2}{4\left[(\Lambda_0^3)^2\cos^2\theta_0-(\Lambda_0^2)^2\right]}.\ea

Now, we are interested in obtaining the explicit expressions for
the conserved charges, which are given by \ba\label{cmom} P_\mu =
\frac{\Lambda^\nu_0}{2\lambda^{0}}\int\int d\xi^1 d\xi^2
g_{\mu\nu},\h \mu,\nu=0,2,3,4.\ea For the present case $P_4=0$
because $X^4=\varphi_3$ does not depend on $\xi^0$. The
computation of the other three conserved quantities leads to the
following expressions for them \ba\label{CE}
&&\frac{E}{\Lambda_0^0}= \frac{2^5\pi
T_2(l_pR)^3\Lambda_1^4\sin\theta_0}
{\left[(\Lambda_0^3)^2\cos^2\theta_0-(\Lambda_0^2)^2\right]^{1/2}}
\ln\left(\frac{1+M}{1-M}\right), \\
\label{CJ1} &&\frac{J_1}{\Lambda_0^2}= \frac{2^7\pi
T_2(l_pR)^3\Lambda_1^4\sin\theta_0}
{\left[(\Lambda_0^3)^2\cos^2\theta_0-(\Lambda_0^2)^2\right]^{1/2}}
\left[\frac{1-M^2}{2}\ln\left(\frac{1+M}{1-M}\right)+M\right], \\
\label{CJ2} &&\frac{J_2}{\Lambda_0^3\cos^2\theta_0}= \frac{2^7\pi
T_2(l_pR)^3\Lambda_1^4\sin\theta_0}
{\left[(\Lambda_0^3)^2\cos^2\theta_0-(\Lambda_0^2)^2\right]^{1/2}}
\left[\frac{1+M^2}{2}\ln\left(\frac{1+M}{1-M}\right)-M\right].\ea

Our next task is to consider the limit, when $M$ tends to its
maximum from below: $M\to 1_-$. In this case, by using (\ref{M^2})
and (\ref{CE})-(\ref{CJ2}), one arrives at the energy-charge
relation \ba\label{2sgm} E-\frac{J_2}{2\cos\theta_0}=\frac{1}{2}
\sqrt{J_1^2+\left[2^7\pi
T_2(l_pR)^3\Lambda_1^4\right]^2\sin^2\theta_0},\ea which is of the
type, expected from gauge theory side.

It is clear that the subtle limit we used to obtain this
dispersion relation shares the features of string derivation.
Namely, while both, the energy and the momentum $J_2$ are
infinite, their difference is finite giving rise to the
magnon-like dispersion relations. Since our formalism is, although
equivalent, somehow different  in notations  and the way of
embedding, one may ask whether one can compare this result with
the ones obtained in string theory. To facilitate the comparison
we will use in the next section the same technique and the same
notations to reproduce well know string results and compare them
to ours.

Other new solutions for M2-branes living in $AdS_7\times S^4$ with
different semiclassical behavior are obtained in the Appendix.

\setcounter{equation}{0}
\section{Comparison with strings on $AdS_5\times S^5$}
For correspondence with the membrane formulae, we will use the
Polyakov action and constraints in diagonal worldsheet gauge
\ba\nl &&S_{gf}= \int
d^{2}\xi\frac{1}{4\lambda^0}\Bigl[G_{00}-\left(2\lambda^0T\right)^2
G_{11}\Bigr],
\\ \nl &&G_{00}+\left(2\lambda^0T\right)^2 G_{11}=0,
\\ \nl &&G_{01}=0,\ea where
\ba\nl G_{mn}= g_{MN}\p_m X^M\p_n X^N,\h \p_m=\p/\p\xi^m, \h m =
(0,1),\h M = (0,1,\ldots,9).\ea The usually used conformal gauge
corresponds to $2\lambda^0T=1$.

We parameterize the metric on $AdS_5\times S^5$ as follows \ba\nl
&&ds^2_{AdS_5}=R^2\left[-\cosh^2\rho dt^2+d\rho^2+\sinh^2\rho
d\Omega^2_3\right],\\ \nl &&ds^2_{S^5}=R^2\left[d\psi^2+\cos^2\psi
d\varphi^2_1 +\sin^2\psi\left(d\theta^2+\cos^2\theta d\varphi^2_2
+\sin^2\theta d\varphi^2_3\right)\right].\ea We will consider
three examples of rotating strings moving in three different
subspaces of the above background.

First, we fix $\rho=0$, $\theta=\theta_0$, $\varphi_3=0$, and
embed the string as follows \ba\nl &&X^0(\xi^m)\equiv t(\xi^m)=
\La_0^0\xi^0, \h X^1(\xi^m)=\psi(\xi^1),
\\ \nl &&X^2(\xi^m)\equiv \varphi_1(\xi^m)=
\La_0^2\xi^0,
\\ \nl &&X^3(\xi^m)\equiv \varphi_2(\xi^m)=
\La_0^3\xi^0,\h \La_0^0,\La_0^2,\La_0^3=constants.\ea This ansatz
corresponds to string extended in the $\psi$- direction, moving
with constant energy $E$ along the $t$-coordinate, and rotating in
the planes defined by the angles $\varphi_1$, $\varphi_2$, with
constant angular momenta $J_1$, $J_2$. Calculations show that in
the limit \ba\nl E\to\infty,\h J_2\to\infty,\h E-J_2 -
\mbox{finite},\h J_1 - \mbox{finite},\ea the above string
configuration is characterized by the following energy-charge
relation \ba\nl E-\frac{J_2}{\cos\theta_0}= \sqrt{J_1^2+\left(4
TR^2\right)^2}.\ea This result reproduces the angular dependence
on the left hand side of (\ref{2sgm}).

As second example, let us fix $\rho=0$, $\theta=\theta_0$,
$\varphi_2=0$, and use the ansatz \ba\nl &&X^0(\xi^m)\equiv
t(\xi^m)= \La_0^0\xi^0, \h X^1(\xi^m)=\psi(\xi^1),
\\ \nl &&X^2(\xi^m)\equiv \varphi_1(\xi^m)=
\La_0^2\xi^0,
\\ \nl &&X^3(\xi^m)\equiv \varphi_3(\xi^m)=
\La_0^3\xi^0.\ea This string configuration is of the same type as
the one just considered. In the above mentioned limit one finds
\ba\nl E-\frac{J_2}{\sin\theta_0}= \sqrt{J_1^2+\left(4
TR^2\right)^2}.\ea

For our third example, we fix $\rho=0$, $\psi=\psi_0$,
$\varphi_1=0$, and choose the string embedding \ba\nl
&&X^0(\xi^m)\equiv t(\xi^m)= \La_0^0\xi^0, \h
X^1(\xi^m)=\theta(\xi^1),
\\ \nl &&X^2(\xi^m)\equiv \varphi_2(\xi^m)=
\La_0^2\xi^0,
\\ \nl &&X^3(\xi^m)\equiv \varphi_3(\xi^m)=
\La_0^3\xi^0,\ea which is of the same type as the previous ones.
Now, in the limit already pointed out, one receives energy-charge
relation given by \ba\nl E-\frac{J_2}{\sin\psi_0}=
\sqrt{\left(\frac{J_1}{\sin\psi_0}\right)^2+\left(4
TR^2\right)^2\sin^2\psi_0}.\ea This reproduces the angular
dependence on the right hand side of (\ref{2sgm}).

Let us mention that in the particular cases when $\cos\theta_0=1$,
or $\sin\theta_0=1$, or $\sin\psi_0=1$, the obtained dispersion
relations reduce to one single formula \ba\nl E-J_2=
\sqrt{J_1^2+\frac{4 \lambda}{\pi^2}},\ea where we have taken into
account that $TR^2=\sqrt{\lambda}/2\pi$. This exactly reproduces
the relation (2.33) of \cite{MTT}. That is why, our three
energy-charge relations represent three different generalizations
of the result given in (2.33) of \cite{MTT} for arbitrary values
of the angles $\theta_0$ and $\psi_0$.

\setcounter{equation}{0}
\section{Concluding remarks}
In this paper we studied particular membrane solutions analogous
to the giant string magnons. The solution we found is supposed to
cover magnon part of the spectrum of gauge spin chain. This
interpretation can be think of in the light of M-theory $\to$
string theory reduction due to the effective $R\times S^1$
topology of our considerations, which make sense in the heavy
brane limit\footnote{We thank A.A. Tseytlin for comments on this
point.}. The membrane configuration we consider develops spikes on
one of the embedding coordinates while on the other depends
linearly on membrane worldvolume coordinates.
Roughly, this can be though as a continuous family of arcs
parameterized by the additional spacial coordinate.
Although there is no concrete
membrane spin chain our solutions to map to, it was interesting to
see whether one can relate the resulting dispersion relations
directly to the gauge theory spin chain. Interestingly, we were
able to obtain membrane configuration that has analogous to the
spin chain dispersion relation. Our result supports the M/gauge
theory correspondence. It also rises the question whether there
exists a spin chain/ladder to which M-theory can be mapped.

One of interesting directions of development is to look for
multi-spin solutions and whether these solutions have the
properties analogous to the corresponding string solutions.
The example considered in this paper seems to be a particular case
of solutions giving rise to such kind of dispersion relations.
Other particular or general solutions would shed more light on
M/gauge theory duality.
It would be interesting also to look for magnon type solutions in the
AdS part of geometry. We hope to report on these issues in the
near future.

\vspace*{.5cm}

{\bf Acknowledgments} \vspace*{.2cm}

This work is supported by NSFB grant under contract $\Phi1412/04$ and Bulgarian NSF BUF-14/06 grant.
We thank A.A. Tseytlin for comments and correspondence.

\newpage
\appendix

\setcounter{equation}{0}
\section{Exact membrane solutions on $AdS_7\times S^4$ and their semiclassical behavior}

\subsection{First type of membrane embedding}

Here, we will use the following coordinates for the background
$AdS_7\times S^4$ metric \ba\nl l_p^{-2}ds^2_{AdS_{7}\times S^4}
&=& 4R^2\left\{-\cosh^2\rho dt^2 + d\rho^2 +
\sinh^2\rho\left(d\psi_1^2 + \cos^2\psi_1 d\psi_2^2+\sin^2\psi_1
d\Omega^2_3\right)\right.\\ \label{AdS7S4m}
&+&\left.\frac{1}{4}\left[d\alpha^2 + \cos^2\alpha
d\theta^2+\sin^2\alpha\left(d\beta^2 + \cos^2\beta d\phi^2
\right)\right]\right\},\\ \nl d\Omega^2_3 &=& d\psi_3^2 +
\cos^2\psi_3 d\psi_4^2 + \cos^2\psi_3\cos^2\psi_4 d\psi_5^2,\h
R^3=\pi N.\ea If we fix \ba\nl \psi_1=\pi/4,\h
\psi_3=\psi_4=\beta=\phi=0,\ea (\ref{AdS7S4m}) reduces to
\ba\label{orb2} ds^2 = (2l_p R)^2\left[-\cosh^2\rho dt^2 + d\rho^2
+ \frac{1}{2}\sinh^2\rho\left(d\psi_2^2+d\psi^2_5\right)
+\frac{1}{4}\left(d\alpha^2 + \cos^2\alpha
d\theta^2\right)\right].\ea

Let us consider the M2-brane embedding $x^M=X^M(\xi^m)$ into
(\ref{orb2}) given by\footnote{The background 3-form on $S^4$ is
zero for this ansatz.} \ba\nl &&X^0(\xi^m)\equiv t(\xi^m)=
\La_0^0\xi^0, \h X^1(\xi^m)=\rho(\xi^2),
\\ \label{la3} &&X^2(\xi^m)\equiv \psi_2(\xi^m)=
\La_0^2\xi^0+\La_1^2\xi^1+\La_2^2\xi^2, \\ \nl &&X^3(\xi^m)\equiv
\psi_5(\xi^m)=
\La_0^3\xi^0-\frac{\La_0^2}{\La_0^3}(\La_1^2\xi^1+\La_2^2\xi^2),
\\ \nl &&X^4(\xi^m)=\alpha(\xi^2),\h X^5(\xi^m)\equiv \theta(\xi^m)=\La_0^5\xi^0,
\h \La_0^0,\ldots,\La_0^5=constants.\ea It corresponds to membrane
extended in the $\rho$- and $\alpha$- directions, moving with
constant energy $E$ along the $t$-coordinate, rotating in the
planes defined by the angles $\psi_2$, $\psi_5$, $\theta$, with
constant angular momenta $S_1$, $S_2$, $J$, and also wrapped along
$\psi_2$, $\psi_5$. The ansatz (\ref{la3}) is a generalization of
the M2-brane embedding considered in \cite{10}. For $\alpha=0$,
one obtains the background felt by the membrane configuration
considered there.

The relation between the parameters in (\ref{la3}) ensures that
the constraints (\ref{0igf}) are identically satisfied. Therefore,
we have to solve the equations of motion (\ref{eqm}) and the
remaining constraint (\ref{00gf}). On the ansatz (\ref{la3}), they
read (the prime is used for $d/d\xi^2$): \ba\nl
&&4\left[K^2(\rho)\rho'\right]'
-\frac{1}{2}\frac{dK^2(\rho)}{d\rho}\left(4\rho'^{2}
+\alpha'^{2}\right)-\frac{1}{2}\frac{\p V(\rho,\alpha)}{\p\rho}=0,
\\ \label{a1} &&\left[K^2(\rho)\alpha'\right]'-\frac{1}{2}\frac{\p V(\rho,\alpha)}{\p\alpha}=0,
\\ \nl &&K^2(\rho)\left(4\rho'^{2}+\alpha'^{2}\right)-V(\rho,\alpha)=0,\ea
where \ba\nl &&K^2(\rho)=8(\lambda^0 T_2)^2(l_p R)^4
\left[1+\left(\Lambda_0^2/\Lambda_0^3\right)^2\right](\Lambda_1^2)^2
\sinh^2\rho, \\ \nl &&V(\rho,\alpha)=(2l_p
R)^2\left\{(\Lambda_0^0)^2\cosh^2\rho -
\frac{1}{2}\left[1+\left(\Lambda_0^2/\Lambda_0^3\right)^2\right](\Lambda_0^3)^2
\sinh^2\rho-\frac{1}{4}(\Lambda_0^5)^2\cos^2\alpha\right\} .\ea

We have not been able to find exact analytical solution of the
above system of nonlinear PDEs. That is why, we restrict ourselves
to the particular case $\La_0^5=0$, i.e. $\theta=0$ and $J=0$.
Thus, the equations (\ref{a1}) reduce to
\ba\label{ras}\rho'=\frac{1}{2K^2(\rho)}\sqrt{K^2(\rho)V(\rho)-A^2},
\h \alpha'=\frac{A}{K^2(\rho)},\h A=const,\ea from where one gets
the equation for the membrane trajectory $\rho(\alpha)$
\ba\label{mt}\frac{d\rho}{d\alpha}=
\frac{1}{2A}\sqrt{K^2(\rho)V(\rho)-A^2}.\ea Here, we are
interested in those solutions of (\ref{ras}) and (\ref{mt}), which
correspond to closed trajectories (orbiting membrane). For them,
$\rho\in (\rho_{min},\rho_{max})$, where $\rho_{min}\equiv \rho_-$
and $\rho_{max}\equiv \rho_+$ are solutions of the equation
$K^2(\rho)V(\rho)-A^2 =0$. In the case under consideration, one
obtains \ba\nl &&x_\pm\equiv\cosh2\rho_\pm
=1+\frac{a^2}{b^2-a^2}\left[1\pm\sqrt{1-\left(\frac{2A}{ac}\right)^2
\frac{b^2-a^2}{a^2}}\right]\ge 1, \\ \nl &&b^2-a^2 > 0, \h
A^2<\left(\frac{ac}{2}\right)^2\frac{a^2}{b^2-a^2},\ea where
\ba\label{abc}&&a^2=(2l_p R)^2(\Lambda_0^0)^2,\h b^2=\frac{1}{2}
(2l_p R)^2
\left[1+\left(\Lambda_0^2/\Lambda_0^3\right)^2\right](\Lambda_0^3)^2,
\\ \nl &&c^2=8(\lambda^0 T_2)^2(l_p R)^4
\left[1+\left(\Lambda_0^2/\Lambda_0^3\right)^2\right](\Lambda_1^2)^2
.\ea

The solution for the orbit is \ba\nl
x(\alpha)\equiv\cosh2\rho(\alpha)=1+\frac{x_--1}
{1-\frac{x_+-x_-}{x_+-1}sn^2\left(\frac{c}{4A}\sqrt{(b^2-a^2)(x_+-1)(x_-+1)}\alpha\right)},\ea
where $sn(u)$ is one of the Jacobian elliptic functions. For the
solutions of the equations (\ref{ras}), one receives \ba\nl
&&\xi^2(x)=\frac{2c(x_--1)}{\sqrt{(b^2-a^2)(x_+-1)(x_-+1)}}
\Pi\left(\varphi_1,\nu,k\right),
\\ \nl &&\xi^2(\alpha)=\frac{2c(x_--1)}{\sqrt{(b^2-a^2)(x_+-1)(x_-+1)}}
\Pi\left(\varphi_2,-\nu,k\right),\ea where $\Pi(\varphi,\nu,k)$ is
one of the elliptic integrals and \ba\nl
&&\varphi_1=\arcsin\sqrt{\frac{(x_+-1)(x-x_-)}{(x_+-x_-)(x-1)}},
\h\varphi_2=am\left(\frac{c}{4A}\sqrt{(b^2-a^2)(x_+-1)(x_-+1)}\alpha\right),
\\ \nl &&\nu=\frac{x_+-x_-}{x_+-1}, \h
k=\sqrt{\frac{2(x_+-x_-)}{(x_+-1)(x_-+1)}}.\ea

The normalization condition \ba\nl 2\pi=\int_{0}^{2\pi}d\xi^2
=2\int_{\rho_{min}}^{\rho_{max}}\frac{d\rho}{\rho'}
=4\int_{\rho_{min}}^{\rho_{max}}\frac{K^2(\rho)d\rho}{\sqrt{K^2(\rho)V(\rho)-A^2}}\ea
is given by \ba\label{cct}
&&\frac{1}{\pi}\sqrt{\frac{c^2}{b^2-a^2}}\int_{x_-}^{x_+}
\sqrt{\frac{x-1}{(x_+-x)(x-x_-)(x+1)}}dx=
\\ \nl &&\sqrt{\frac{c^2}{b^2-a^2}}\sqrt{\frac{x_--1}{x_-+1}}F_{1}
\left(1/2,1/2,-1/2;1;-\frac{x_+ - x_-}{x_-+1},-\frac{x_+ -
x_-}{x_--1}\right)= \\ \nl
&&\sqrt{\frac{c^2}{b^2-a^2}}\sqrt{\frac{x_--1}{x_-+1}}
\left(1+\frac{x_+ - x_-}{x_-+1}\right)^{-1/2} \left(1+\frac{x_+ -
x_-}{x_--1}\right)^{1/2} \times \\ \nl &&F_{1}
\left(1/2,1/2,-1/2;1;\frac{1}{1+\frac{x_-+1 }{x_+
-x_-}},\frac{1}{1+\frac{x_--1}{x_+ - x_-}}\right)= 1,\ea where
$F_1(a,b_1,b_2;c;z_1,z_2)$ is one of the hypergeometric functions
of two variables \cite{PBM-III}.

For the conserved quantities $E$, $S_1$ and $S_2$, on the above
membrane solution,  we obtain the following expressions
\ba\label{cE}
&&\frac{E}{\Lambda_0^0}=\frac{(2l_pR)^2}{2\lambda^0}\int\int
d\xi^1d\xi^2\cosh^2\rho(\xi^2)=\\ \nl
&&\frac{(2l_pR)^2\pi^2c}{2\lambda^0}\sqrt{\frac{(x_-+1)(x_--1)}{b^2-a^2}}
F_{1} \left(1/2,-1/2,-1/2;1;-\frac{x_+ - x_-}{x_-+1},-\frac{x_+ -
x_-}{x_--1}\right)= \\ \nl
&&\frac{(2l_pR)^2\pi^2c}{2\lambda^0}\sqrt{\frac{(x_-+1)(x_--1)}{b^2-a^2}}
\left(1+\frac{x_+ - x_-}{x_-+1}\right)^{1/2} \left(1+\frac{x_+ -
x_-}{x_--1}\right)^{1/2} \times \\ \nl &&F_{1}
\left(1/2,-1/2,-1/2;1;\frac{1}{1+\frac{x_-+1 }{x_+
-x_-}},\frac{1}{1+\frac{x_--1}{x_+ - x_-}}\right),\ea
\ba\label{cS}
&&\frac{S_1}{\Lambda_0^2}=\frac{S_2}{\Lambda_0^3}=\frac{(2l_pR)^2}{4\lambda^0}\int\int
d\xi^1d\xi^2\sinh^2\rho(\xi^2)=\\ \nl
&&\frac{(2l_pR)^2\pi^2c}{4\lambda^0}\sqrt{\frac{(x_--1)^3}{(b^2-a^2)(x_-+1)}}
F_{1} \left(1/2,1/2,-3/2;1;-\frac{x_+ - x_-}{x_-+1},-\frac{x_+ -
x_-}{x_--1}\right)= \\ \nl
&&\frac{(2l_pR)^2\pi^2c}{4\lambda^0}\sqrt{\frac{(x_--1)^3}{(b^2-a^2)(x_-+1)}}
\left(1+\frac{x_+ - x_-}{x_-+1}\right)^{-1/2} \left(1+\frac{x_+ -
x_-}{x_--1}\right)^{3/2} \times \\ \nl &&F_{1}
\left(1/2,1/2,-3/2;1;\frac{1}{1+\frac{x_-+1 }{x_+
-x_-}},\frac{1}{1+\frac{x_--1}{x_+ - x_-}}\right).\ea

In the semiclassical limit (large conserved charges), which in the
present case correspond to \ba\nl x_+\to
\frac{2a^2}{b^2-a^2}\to\infty,\h x_-\to 1,\ea the equalities
(\ref{cct}), (\ref{cE}) and (\ref{cS}) simplify to \ba\nl
c^2=b^2-a^2,\h \frac{E}{\Lambda_0^0}=\frac{(2\pi
l_pR)^2a^2}{2\lambda^0(b^2-a^2)},\h
\frac{S_1}{\Lambda_0^2}=\frac{S_2}{\Lambda_0^3}=\frac{(2\pi
l_pR)^2a^2}{4\lambda^0(b^2-a^2)}.\ea From here, one receives the
following energy-charge relation \ba\label{ecr1}
\left[2(S_1^2+S_2^2)-E^2\right]^3 -
8(l_pR)^6(\pi^2T_2\Lambda_1^2)^2\left(1+\frac{S_1^2}{S_2^2}\right)E^4=0,\ea
where $\Lambda_1^2$ is wrapping parameter with integer values (see
(\ref{la3})). (\ref{ecr1}) is third order algebraic equation for
$E^2$, and does not give a 2-spin generalization of the known
$E-S\sim S^{1/3}$ relation for membranes in $AdS_p\times S^q$
backgrounds. If we set $\Lambda_1^2=0$, this will correspond to
string-like ansatz for the membrane, because the $\xi^1$
worldvolume coordinate drops out from the embedding. Then we get
\ba\nl E=\sqrt{2(S_1^2+S_2^2)},\ea which has nothing to do with
the 2-spin giant magnon dispersion relations. For say $S_1=0$, it
reduces to $E-\sqrt{2}S_2=0$, which should be compared with the
spin chain ferromagnetic vacuum characterized by the relation
$E-J=0$. For $S_1<<S_2$, one can rewrite it as \ba\nl
E-\sqrt{2}S_2=\frac{S_1^2}{\sqrt{2}S_2}+\ldots .\ea Obviously,
this relation does not correspond to the low energy spin waves
states of the spin chain, dual to spinning strings with \ba\nl
E-J\sim \frac{\lambda}{J}+\ldots .\ea The conclusion is that even
this stringy-restricted membrane configuration does not describe
any part of the exited spin chain spectrum.

Let us finally recall that the above results are obtained in the
framework of the particular case $\theta=0$, when the background
seen by the membrane is \ba\label{orb3} ds^2 = (2l_p
R)^2\left[-\cosh^2\rho dt^2 + d\rho^2 +
\frac{1}{2}\sinh^2\rho\left(d\psi_2^2+d\psi^2_5\right)
+\frac{1}{4}d\alpha^2\right],\ea for which the Lagrangian density
in the action (\ref{omagf}), on the ansatz (\ref{la3}) (for
$\Lambda_0^5=0$), reduces to \ba\label{rma}
\mathcal{L}=-\frac{1}{4\lambda^0}\left[K^2(\rho)\left(4\rho'^{2}+\alpha'^{2}\right)
+V(\rho)\right].\ea

\subsection{Second type of membrane embedding}
Here , we intend to consider the membrane configuration described
in the previous subsection from another viewpoint, by representing
it as embedded in flat space-time. To this end, we rewrite the
metric (\ref{im}) induced on the membrane worldvolume in the form
\ba\nl G_{mn}=\eta_{MN}\p_{(m}Z^M\p_{n)}\bar{Z}^N=
\eta_{\alpha\beta}\p_{(m}Y^\alpha\p_{n)}\bar{Y}^\beta
+\p_{(m}X\p_{n)}\bar{X},\h \eta_{\alpha\beta}=diag(-1,1,1),\ea
where $Y^\alpha$ and $X$ are complex coordinates given by
\ba\label{fc} &&Y^0=2l_pR e^{it}\cosh\rho,\h Y^1=\sqrt{2}l_pR
e^{i\psi_2}\sinh\rho,\\ \nl &&Y^2=\sqrt{2}l_pR
e^{i\psi_5}\sinh\rho,\h X=l_pR e^{i\alpha},\ea and satisfying the
equalities \ba\label{fscs}\eta_{\alpha\beta}Y^\alpha\bar{Y}^\beta=
-(2l_pR)^2,\h X\bar{X}=(l_pR)^2.\ea In the coordinates (\ref{fc}),
the flat target space metric \ba\nl ds^2_f
=\eta_{\alpha\beta}dY^\alpha d\bar{Y}^\beta +dXd\bar{X}\ea
coincides with the one in (\ref{orb3}). In order to reproduce the
reduced Lagrangian density (\ref{rma}) corresponding to the
membrane solution obtained in the previous subsection, one just
have to replace $t$, $\rho$, $\psi_2$, $\psi_5$ and $\alpha$ in
(\ref{fc}) with the expressions for them given in (\ref{la3}).

Now, let us consider the membrane configuration given by the
embedding \ba\nl &&y^0(\xi^0,\xi^2)=2l_pR
e^{i\La_0^0\xi^0}r_1(\xi^2),\h y^1(\xi^m)=\sqrt{2}l_pR
e^{i(\La_0^2\xi^0+\La_1^2\xi^1+\La_2^2\xi^2)}r_2(\xi^2),\\
\label{fse} &&y^2(\xi^m)=\sqrt{2}l_pR
e^{i\left[\La_0^3\xi^0-\frac{\La_0^2}{\La_0^3}(\La_1^2\xi^1+\La_2^2\xi^2)\right]
}r_2(\xi^2),\h y^3(\xi^2)=l_pR e^{i\alpha(\xi^2)},\ea where in
accordance with (\ref{fscs}), $r_1$ and $r_2$ are constrained
by\footnote{The second constraint in (\ref{fscs}) is identically
satisfied for this ansatz.} \ba\label{r12} r_1^2-r_2^2-1=0.\ea The
corresponding reduced Lagrangian density is
\ba\label{frl}\mathcal{L}_f=
\frac{1}{4\lambda^0}\left\{c^2r_2^2\left[4(r'^{2}_{1}-r'^{2}_{2})-\alpha'^{2}\right]
-a^2r_1^2+b^2r_2^2\right\}-\Lambda(r_1^2-r_2^2-1).\ea Here, the
constant coefficients $a^2$, $b^2$, $c^2$ are introduced in
(\ref{abc}), and $\Lambda$ is Lagrange multiplier. For the
embedding (\ref{fse}), the constraints (\ref{0igf}) are
identically satisfied. The remaining constraint (\ref{00gf}) takes
the form
\ba\label{00f}c^2r_2^2\left[4(r'^{2}_{1}-r'^{2}_{2})-\alpha'^{2}\right]
+a^2r_1^2-b^2r_2^2=0.\ea

From (\ref{frl}), one obtains the following equations of
motion\footnote{At this stage, we fix $\Lambda=constant$.}
\ba\label{er1}
&&(r_2^2r'_1)'=-\frac{a^2+4\lambda^0\Lambda}{4c^2}r_1, \\ \nl
&&(r_2^2r'_2)'= -\left[\frac{b^2+4\lambda^0\Lambda}{4c^2}
+r'^{2}_{1}-r'^{2}_{2}
-\frac{1}{4}\alpha'^{2}\right]r_2, \\
\nl &&(r_2^2\alpha')'=0, \h r_1^2-r_2^2-1=0.\ea The solution of
the equation for $\alpha'$ is \ba\label{sa'}
\alpha'=\frac{2C_\alpha}{r_2^2},\h C_\alpha=const.\ea The
replacement of (\ref{sa'}) into the equation for $r'_2$ and
(\ref{00f}) gives \ba\label{er2}&&(r_2^2r'_2)'=
-\left[\frac{b^2+4\lambda^0\Lambda}{4c^2} +r'^{2}_{1}-r'^{2}_{2}
-\frac{C_\alpha^2}{r_2^4}\right]r_2, \\ \label{00fr}
&&r'^{2}_{1}-r'^{2}_{2}=\frac{C_\alpha^2}{r_2^4} -
\frac{1}{4c^2}\left(a^2\frac{r_1^2}{r_2^2}-b^2\right).\ea With the
help of (\ref{00fr}) and $r_1^2-r_2^2-1=0$, the equation
(\ref{er2}) reduces to \ba\label{r2'} r'_2=\frac{1}{2\sqrt{2}c
r_2^2}
\sqrt{-(2b^2-a^2+4\lambda^0\Lambda)r_2^4+2a^2r_2^2+C_{r_2}},\ea
where $C_{r_2}$ is arbitrary integration constant. By using the
equality $r_1^2-r_2^2=1$ once again, this time in (\ref{er1}), one
derives for $r'_1$ the expression \ba\label{r1'}
r'_1=\frac{1}{2\sqrt{2}c (r_1^2-1)}
\sqrt{-(a^2+4\lambda^0\Lambda)\left[r_1^2(r_1^2-2)+C_{r_1}\right]},
\h C_{r_1}=const.\ea In order (\ref{00fr}) to be identically
satisfied on the solutions (\ref{r2'}), (\ref{r1'}), the
integration constants must be related as follows \ba\nl
8c^2C_\alpha^2+C_{r_1}+C_{r_2}=a^2+4\lambda^0\Lambda.\ea Of
course, the condition $r_1^2-r_2^2=1$ will give another relation
between $C_\alpha$, $C_{r_1}$ and $C_{r_2}$.

For our further purposes, it is enough to solve one of the
differential equations (\ref{r2'}), (\ref{r1'}), and we choose the
first one. Setting $r'_2=0$, we observe that there exist two
different cases, which correspond to periodic motion, depending on
the sign of $C_{r_2}$.
\paragraph{First case ($y\equiv r_2^2$)}
\ba\nl &&y_1=-\frac{B^2}{\mid A\mid}\left(\sqrt{1+\frac{2\mid
A\mid\mid C\mid}{B^4}}-1\right)<0,\\ \nl &&y_{max}=y_2
=\frac{B^2}{\mid A\mid}\left(1+\sqrt{1+\frac{2\mid A\mid\mid
C\mid}{B^4}}\right),\h\mbox{therefore}\h y_{min}=0.\ea
\paragraph{Second case}
\ba\nl &&y_{min}=y_1=\frac{B^2}{\mid
A\mid}\left(1-\sqrt{1-\frac{2\mid A\mid\mid
C\mid}{B^4}}\right)>0,\\ \nl &&y_{max}=y_2 =\frac{B^2}{\mid
A\mid}\left(1+\sqrt{1-\frac{2\mid A\mid\mid
C\mid}{B^4}}\right)>y_{min}.\ea For convenience, we have
introduced here the notations \ba\nl
A=-\frac{2b^2-a^2+4\lambda^0\Lambda}{4c^2}<0,\h
B^2=\frac{a^2}{4c^2},\h C=\frac{C_{r_2}}{8c^2}.\ea

We begin with considering the first case, for which the solution
of (\ref{r2'}) is \ba\nl
\xi^2(y)=\frac{1}{3}\left(\frac{2y^3}{\mid A\mid\mid y_1\mid
y_{max}}\right)^{1/2} F_1\left(3/2,1/2,1/2;5/2; -\frac{y}{\mid
y_1\mid},\frac{y}{y_{max}}\right).\ea The normalization condition
\ba\nl
2\pi=\int_{0}^{2\pi}d\xi^2=2\int_{(r_2)_{min}}^{(r_2)_{max}}\frac{dr_2}{r'_2},\ea
leads to \ba\label{ncc1} \frac{y_{max}}{\left(2^3\mid A\mid\mid
y_1\mid\right)^{1/2}}\mbox{}_2F_{1}\left(3/2,1/2;2;-\frac{y_{max}}{\mid
y_1\mid}\right)=1.\ea For the conserved quantities $E$, $S_1$ and
$S_2$, canonically conjugated to the coordinates $t$, $\psi_2$ and
$\psi_5$, we obtain the expressions \ba\nl
&&\frac{E}{\Lambda_0^0}=\frac{\pi}{2\lambda^0}(2l_pR)^2(I+2\pi),
\h \frac{S_1}{\Lambda_0^2}=\frac{S_2}{\Lambda_0^3}=
\frac{\pi}{4\lambda^0}(2l_pR)^2I, \\ \label{ccc1} &&I=\frac{3\pi
y_{max}^2}{\left(2^5\mid A\mid\mid
y_1\mid\right)^{1/2}}\mbox{}_2F_{1}\left(5/2,1/2;3;-\frac{y_{max}}{\mid
y_1\mid}\right).\ea

In the semiclassical limit, which in the case under consideration
correspond to \ba\nl y_{max},\mid y_1\mid\to \sqrt{2\frac{\mid
C\mid}{\mid A\mid}}\to\infty,\ea the equalities (\ref{ncc1}) and
(\ref{ccc1}) simplify to \ba\nl \frac{2^5\pi^2\mid
C\mid}{\Gamma^8(1/4)\mid A\mid^3}=1, \h I=\left[\frac{2^3\pi^6
\mid C\mid^{3}}{\Gamma^8(3/4)\mid A\mid^5}\right]^{1/4},\ea from
where one derives the following dependence of the energy $E$ on
the charges $S_1$, $S_2$ \ba\label{ecrc1}
E^2&=&4(S_1^2+S_2^2)-\frac{\Gamma^8(1/4)(l_pR)^2}{2^4\lambda^0}\mid
C_{r_{2}}\mid
\\ \nl &&+\Lambda\left[\frac{\Gamma^{32}(1/4) \mid
C_{r_{2}}\mid^2}{2^{14}\pi^2(\lambda^0)^7(T_2\Lambda_1^2)^4(l_pR)^2}\right]^{1/3}
\left(1+\frac{S_1^2}{S_2^2}\right)^{-2/3}.\ea Here, $\Lambda_1^2$
is a winding number as in the previous solution, while the
integration constant $C_{r_{2}}$ and the Lagrange multipliers
$\lambda^0$, $\Lambda$ are free parameters. This is a
generalization of the energy-charge relation $E-S\sim S^{-1}$,
unknown for M2-branes up to now.

Let us turn to the second case. The solution of (\ref{r2'}) reads
($\Delta y=y-y_{min}$, $\Delta y_m=y_{max}-y_{min}$) \ba\nl
\xi^2(y)=\left(\frac{2y_{min}\Delta y}{\mid A\mid\Delta
y_{m}}\right)^{1/2} F_1\left(1/2,-1/2,1/2;3/2; -\frac{\Delta
y}{y_{min}},\frac{\Delta y}{\Delta y_{m}}\right).\ea The
normalization condition gives \ba\label{ncc2}
\left(\frac{y_{min}}{2\mid A\mid}\right)^{1/2}\left(1+\frac{\Delta
y_m}{y_{min}}\right)^{1/2}\mbox{}_2F_{1}\left(1/2,-1/2;1;\frac{1}{1+\frac{y_{min}}{\Delta
y_{m}}}\right)=1.\ea The explicit expressions for the conserved
quantities can be obtained from (\ref{ccc1}) by the replacement
$I\to J$, with $J$ given by \ba\label{ccc2}\pi
\left(\frac{2y^3_{min}}{\mid
A\mid}\right)^{1/2}\left(1+\frac{\Delta
y_m}{y_{min}}\right)^{3/2}\mbox{}_2F_{1}\left(1/2,-3/2;1;\frac{1}{1+\frac{y_{min}}{\Delta
y_{m}}}\right).\ea

Taking the semiclassical limit, which now corresponds to \ba\nl
y_{max}\to \frac{2B^2}{\mid A\mid}\to\infty,\h y_{min}\to 0,\ea
one gets the following energy charge-relation \ba\label{ecrc2l}
&&\left[1+(2/\pi^2)(2l_pR)^2(2\lambda^0T_2\Lambda_1^2)^2
\left(1+S_1^2/S_2^2\right)\right]E^2 \\
\nl &&-\frac{2^{3/2}\pi^3\Lambda
E}{3l_pR\lambda^0T_2\Lambda_1^2\left(1+S_1^2/S_2^2\right)^{1/2}}-4(S_1^2+S_2^2)=0,\ea
where $E(S_1,S_2)$ is given by the positive root of this equation.

An interesting question is whether the membrane configuration just
considered can be related to the known integrable systems
appearing in the context of rigid string motion on $AdS_5\times
S^5$. To make such comparison, we replace the solution (\ref{sa'})
for $\alpha'$ into (\ref{frl}) and obtain \ba\nl\mathcal{L}=
\frac{1}{4\lambda^0}\left\{4c^2r_2^2(r'^{2}_{1}-r'^{2}_{2})
-a^2r_1^2+b^2r_2^2-\frac{4c^2C_{\alpha}^2}{r_2^2}\right\}-\Lambda(r_1^2-r_2^2-1).\ea
This Lagrangian is similar to a particular case of the Lagrangian
describing the n-dimensional Neumann-Rosochatius-like integrable
system with indefinite metric \cite{ART0311} \ba\nl
\mathcal{L}_{NR-l}= \frac{1}{2}\eta^{rs}\left(r'_r r'_s
-\omega_r^2 r_s r_s - \frac{u_r u_s}{r_r r_s}\right)
-\frac{1}{2}\tilde{\Lambda}(\eta^{rs}r_r r_s+1),\h \eta = diag
(-,+,...,+,-).\ea The essential difference is that in the membrane
case the kinetic term depends on the coordinate $r_2$. However,
this does not exclude the existence of M2-brane configurations
with Lagrangians of the type $\mathcal{L}_{NR-l}$.


\end{document}